\documentclass[epj-spec]{svjour}

\usepackage{graphics}
\usepackage[dvips]{graphicx}
\usepackage{latexsym}
\usepackage{epsfig}
\usepackage{bm}
\usepackage{times,psfrag,subfigure}        
\usepackage{amssymb,amsmath}
\begin{document}

\title{Modelling capillary filling dynamics using lattice Boltzmann simulations}
\author{C. M. Pooley \and H. Kusumaatmaja  \and J. M. Yeomans 
\thanks{\email{j.yeomans1@physics.ox.ac.uk}}}
\institute{The Rudolf Peierls Centre for Theoretical Physics, Oxford University, 1 Keble Road, 
Oxford OX1 3NP, U.K.}
\abstract{
  We investigate the dynamics of capillary filling using two lattice Boltzmann schemes: a 
liquid-gas model and a binary model.
  The simulation results are compared to the well-known Washburn's law, which predicts that 
the filled length of the capillary scales with time as $l \propto t^{1/2}$.
  We find that the liquid-gas model does not reproduce Washburn's law due to condensation of 
the gas phase   at the interface, which causes the asymptotic behaviour of the capillary 
penetration to be faster than   $t^{1/2}$. The binary model, on the other hand, 
  captures the correct scaling behaviour when the viscosity ratio between the two phases is 
sufficiently high.
}

\maketitle

\section{Introduction}

\begin{figure}
\begin{center}
\includegraphics[width = 4.in]{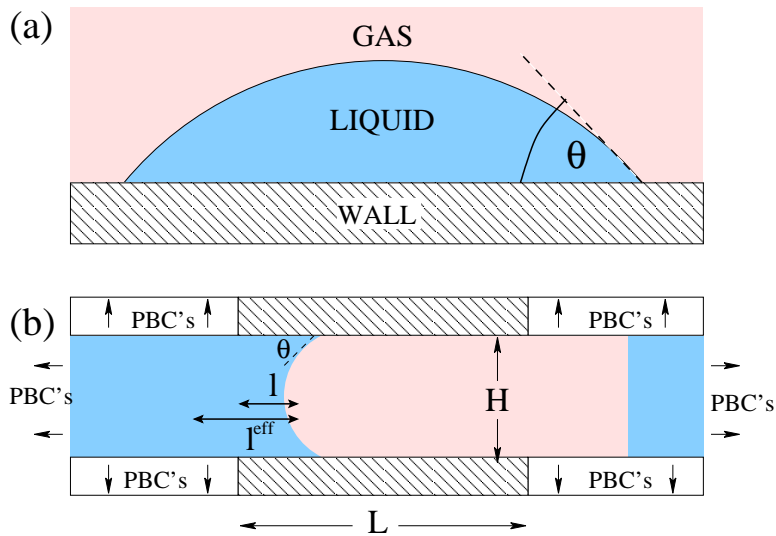}
\end{center}
\caption{The two simulation setups. (a) A drop resting at equilibrium on a surface. ($\theta$ 
is the contact angle.) (b) Capillary flow along a tube. $l$ is the length of tube filled with fluid 
and $l^{eff}=l+H/2$ is an effective filled length.}
\label{fig7}
\end{figure}
%


The aim of this paper is to present an effective modelling method for simulating liquid which 
fills a capillary tube that initially contains gas. Capillary
tubes have a hydrophilic inner surface and the energy liberated in wetting this surface is 
used to drive the fluid up the tube.

The classical analysis of capillary penetration is due to Washburn \cite{Washburn}. He assumed 
that the liquid is incompressible and that the fluid flow  has a parabolic profile (Poiseuille 
flow).
In two dimensions, the average velocity of a parabolic flow is 
\begin{equation} 
\bar{v} = -\frac{H^2}{12 \eta} \frac{dp}{dx} \label{eq1}
\end{equation}
where $H$ is the capillary tube width, $\eta$ is the liquid viscosity, and $dp/dx$ is the 
pressure gradient
that sets up the flow. The Laplace pressure drop across a curved interface of radius $R$ is 
given by $\Delta p = \gamma_{lg}/R$, where $\gamma_{lg}$ is the liquid-gas surface tension. 
The pressure gradient in the fluid is therefore
\begin{equation}
\frac{dp}{dx} = - \frac{\gamma_{lg}}{Rl} \label{eq2}
\end{equation}
where $l$ is the length of the liquid column that has penetrated the capillary. $R$ is related 
to the dynamic contact angle $\theta$ (see Fig. \ref{fig7}(b) for a definition of $\theta$) 
through $R= H/2\cos{\theta}$. By substituting Eq. (\ref{eq2}) into (\ref{eq1}) and using 
$\bar{v} = dl/dt$ we obtain 
\begin{equation}
l = \left(\frac{\gamma_{lg} H \cos{\theta}}{3\eta}\right)^{\tfrac{1}{2}} 
(t+t_0)^{\tfrac{1}{2}} 
\label{eq3}
\end{equation}
where $t_0$ is an integration constant. An alternative way to derive Eq. (\ref{eq3}) is to 
equate the dissipation of energy due to viscosity to the energy liberated as the liquid wets 
the surface. In the original analysis, Washburn neglected the viscous dissipation of energy in 
the gas phase (the viscosity ratio between water and air is around $\sim 10^3$ so this is a 
very good approximation in this case) and the deviation from the Poiseuille flow velocity 
profile at the inlet and near the curved interface \cite{Levine}.

In this paper we consider two approaches to modelling capillary dynamics. The first is a van 
der Waals liquid-gas model and the second is a binary fluid model with a viscosity difference 
between the two phases.  We show that the liquid-gas model does not reproduce Washburn's law 
due to condensation of the gas phase at the interface, which causes the asymptotic behaviour 
of the capillary penetration to be faster than   $t^{1/2}$. The binary model, however,  captures the correct scaling behaviour when the viscosity ratio between the two phases is 
sufficiently high. Other authors  \cite{Timonen,santos,succi1,succi2} have discussed different lattice Boltzmann approaches to model capillary filling.   Their results are broadly similar to those reported here, but the models differ in the details of the behaviour of the dynamic contact line.

\section{A liquid-gas model}

The pressure tensor for a liquid-gas system, resulting from a Landau free energy functional, 
is  \cite{swift,yeomans}
\begin{eqnarray}
  P_{\alpha \beta} =  \left(p_0 - \kappa \rho \nabla^2 \rho - \frac \kappa 2 |\nabla \rho|^2 
\right) \delta_{\alpha \beta} + \kappa \partial_{\alpha} \rho  \, \partial_{\beta} \rho,
\label{eq:pt}
\end{eqnarray}
where
\begin{eqnarray}
p_0 = \frac{\rho T}{1-b \rho} - a\rho^2
\end{eqnarray}
is the van der Waals bulk pressure, $\rho$ is the fluid density and $\kappa$ is a parameter 
related to the surface tension. 
This leads to liquid-gas phase separation below a critical temperature. For this investigation 
we chose the parameters $\kappa = 0.02$, $a = 9/49$, $b = 2/21$ and $T = 0.56$ giving a 
surface tension of $\gamma_{lg} = 0.0112$ and liquid and gas densities of $\rho_l = 4.54$ and 
$\rho_g = 2.59$, respectively.  

\subsection{Lattice Boltzmann implementation}

We used a two-dimensional, nine velocity vector, free energy, multiple-relaxation-timescale 
lattice Boltzmann method \cite{succi}. 
Here we briefly outline the method and refer the reader to \cite{ourpaper} for more details.
The system is divided up into a square grid of nodes, and on each node there is a particle 
distribution function $f_i({\bf r}, t)$. The label $i$ denotes a particular lattice velocity 
${\bf e}_i$, defined by ${\bf e}_0 = (0,0)$, ${\bf e}_{1,2} = (\pm c,0)$, ${\bf e}_{3,4} = 
(0,\pm c)$, ${\bf e}_{5,6} = (\pm c,\pm c)$, and ${\bf e}_{7,8} = (\mp c,\pm c)$. The lattice 
speed $c$ is given by $c = \tfrac{\Delta x}{\Delta t}$.   

The time evolution equation for the particle distribution function is 
\begin{eqnarray}
{\bf f}({\bf r} + {\bf e} \Delta t , t+\Delta t) = {\bf f}({\bf r}, t) - {\bf M}^{-1} {\bf S} 
{\bf M} \left[ {\bf f} - {\bf f}^{eq} \right], 
\label{latbolt}
\end{eqnarray}
where $f_i$ has been written as a column vector, ${\bf M}$ is a matrix that performs a change 
of basis, ${\bf S}$ is a diagonal matrix which defines different relaxation times for 
different modes and ${\bf f}^{eq}$ is an equilibrium distribution function. Details of a 
suitable choice for ${\bf M}$, ${\bf S}$ and ${\bf f}^{eq}$ are given in Appendix \ref{app1}. 

In the limit of long length and timescales, Eq. (\ref{latbolt}) leads to the continuum 
Navier-Stokes equation
\begin{eqnarray}
 \partial_{t}(\rho v_\alpha) +  \partial_\beta ( \rho v_\alpha v_\beta) = -  \partial_\beta
P_{\alpha \beta} + \partial_\beta  \left(  \nu \rho \left[ \partial_\beta  v_\alpha +
\partial_\alpha  v_\beta \right]  \right),
\label{nsfinal2}
\end{eqnarray}
which determines the dynamics of the system. The parameter $\nu$, the kinematic viscosity, is 
related to the dynamic viscosity by $\eta = \rho \nu$.

\subsection{Numerical results}


%
\begin{figure}
\begin{center}
\includegraphics[width = 5.6in]{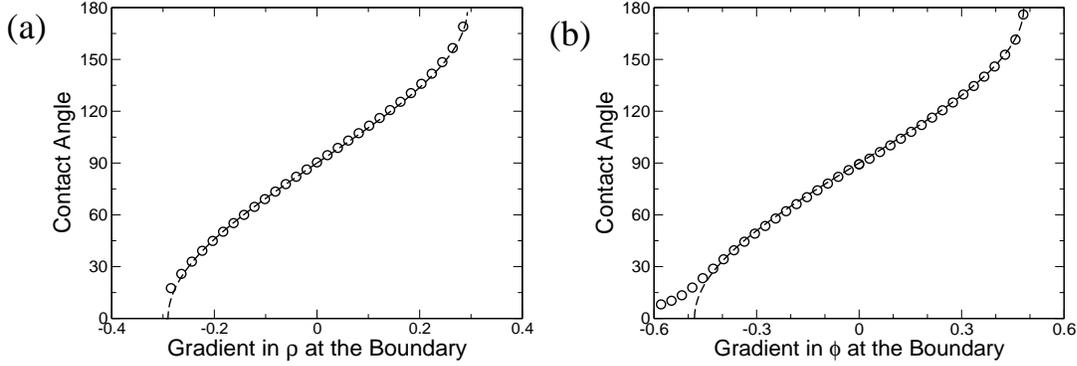}
\end{center}
\caption{ The equilibrium contact angle as a function of the equilibrium gradient in $\rho$ or 
$\phi$ at the boundary for (a) the  
liquid-gas system and (b) the binary system. The circles were obtained numerically by 
quasistatically scanning the equilibrium gradient in time, using $\nu = 1/6$ in (a) and 
$\nu_l=0.83$ and $\nu_g=0.067$ in (b). The dashed curve comes from Young's law (\ref{Young}), 
based on numerically calculated surface tensions.}
\label{fig2}
\end{figure}
To check the equilibrium properties of the model, we numerically measured the contact angle 
and compared it with theory.
Figure \ref{fig7}(a) shows a drop resting on top of a solid surface. In general, the 
solid-liquid $\gamma_{sl}$, solid-gas $\gamma_{sg}$ and liquid-gas $\gamma_{lg}$ surface 
tensions are different. At the contact point $A$ the balance of forces gives Young's law:
\begin{eqnarray}
\cos \theta^{eq} = \frac{\gamma_{sg} - \gamma_{sl}}{\gamma_{lg}},
\label{Young}
\end{eqnarray}
where $\theta^{eq}$ defines the equilibrium contact angle. 

We used a system of size $300\times50$ lattice units and placed an initially semi-circular 
drop on the lower wall. Non-slip boundary conditions were simulated by using a bounce-back 
scheme as well as setting boundary nodes to rest after each streaming step (this and the 
MRT-LB scheme were found to be necessary to stop unphysical currents appearing near to the 
interfaces \cite{ourpaper}). The wetting properties of the surface were changed by altering 
the gradient in $\rho$, $\left. \partial_n \rho \right|_b$, as it appears in the equilibrium 
distribution, at the boundary \cite{Briant1}.  
Figure \ref{fig2}(a) shows that there is good agreement between the numerically measured 
contact angle (the circles) and that predicted by Young's law (the dashed line).

\begin{figure}
  \begin{center}
  \includegraphics[width = 3.in]{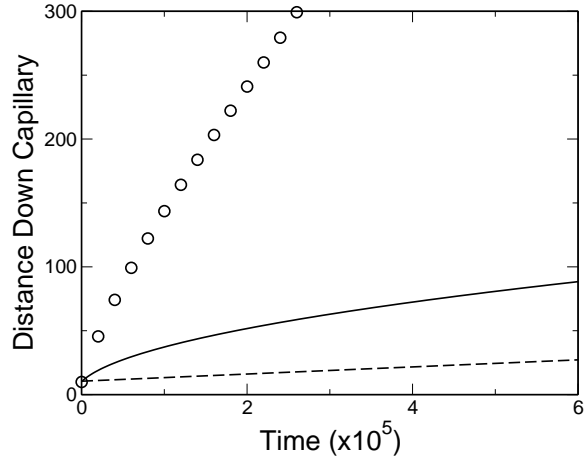}
\end{center}
\caption{ The distance of the liquid-gas interface along the capillary as a function of time. 
Circles
 are simulation results using $\nu = 1/6$, the solid line is Washburn's law (\ref{eq3}) and 
the dashed line is Washburn's law taking into account dissipation in the gas phase.}
\label{fig10}
\end{figure}
Next, we focus on simulating a dynamical system, that of capillary filling, and test the 
Washburn relation in Eq. (\ref{eq3}).
The simulation setup is illustrated in Fig. \ref{fig7}(b). The system consists of a lattice of 
size $700\times40$ lattice units with periodic boundary conditions in the $x$ direction. The 
upper and lower sides of the system have two sets of boundary conditions. In the middle 
portion the boundaries are non-slip and wetting (denoted by the hashed region in the diagram) 
and this represents the sides of the capillary. The length of this is set to be $L = 350$.  
At either side of the capillary the boundary conditions are periodic in the $y$ direction, and 
hence allow slip, and these areas represent a reservoir of liquid and gas. 

Using the results from Fig. \ref{fig2}(a) we chose the equilibrium gradient at the boundary to 
be $\left. \partial_n \rho \right|_b = -0.144$ such that the equilibrium contact angle is 
$\theta^{eq} = 60 ^\circ$. This wetting interaction induces the liquid and gas interface to 
form a meniscus. The system is initialised such that $l=10$ and evolved in time using the 
lattice Boltzmann algorithm. The solid line in Fig. \ref{fig10} shows a plot of $l$ against 
time. When compared against the theoretical prediction of Washburn (given by the dashed line) 
we observe that agreement is poor. One reason behind this might be that viscous dissipation in 
the gas phase cannot be ignored. However, taking this into account (as shown be the dotted 
line) makes the predicted flow even slower.

\begin{figure}
\begin{center}
\includegraphics[width = 5.6in]{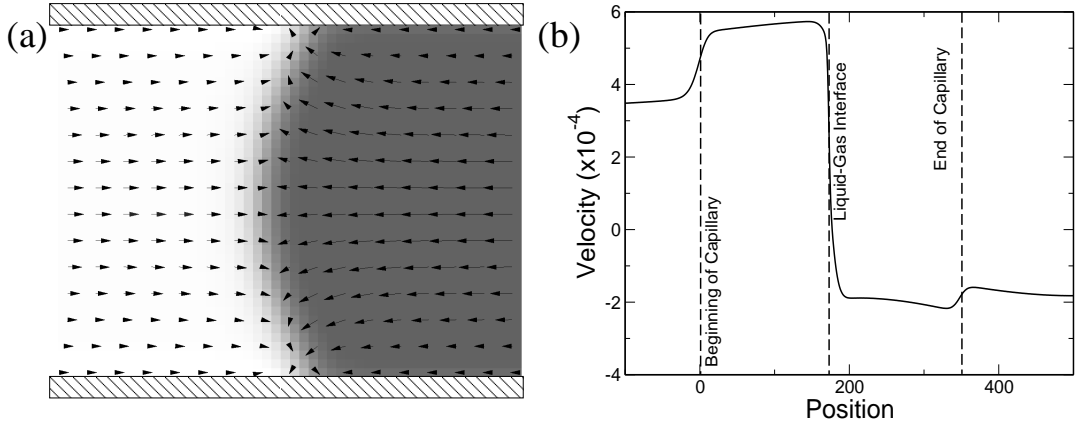}
\end{center}
\caption{(a) The velocity field around the liquid-gas interface during capillary filling using 
the standard liquid-gas model. (b) The $x$ component of the velocity as a function of $x$ at 
the centre of the capillary.}
\label{fig3}
\end{figure}

The reason behind the discrepancy between theory and simulation can be clearly seen if we 
observe the fluid flow profile near to the meniscus, as plotted in Fig. \ref{fig3}(a).
We find that whilst fluid in the liquid phase is moving to the right and filling up the 
capillary, the fluid in the gas phase is moving ${\it left}$ and condensing to form liquid, 
which helps to significantly increase the speed of the interface. This is seen even more 
clearly in  Fig. \ref{fig3}(b), which shows the $x$ component of the fluid velocity as a 
function of distance down the tube.
It is because the system primarily exhibits a condensation driven interface velocity, rather 
than sucking fluid through the capillary, that Washburn's law breaks down.


\section{A binary model}

\subsection{Lattice Boltzmann implementation}

One way to prevent condensation is to simulate the system as a binary fluid \cite{swift,yeomans}. We associate $A$ 
particles with the liquid phase and $B$ particles with the gas phase. Because particle species 
is strictly conserved, condensation is no longer permitted. Obtaining a viscosity ratio 
between the two phases is achieved by making the kinematic viscosity $\nu$ a function of the 
order parameter $\phi = (\rho_A - \rho_B)/(\rho_A + \rho_B)$:
\begin{eqnarray}
\nu = \nu_g + \tfrac{\phi+1}{2} \left(\nu_l-\nu_g \right)
\label{nu}
\end{eqnarray}
such that the viscosity has the bulk values $\nu_l$ and $\nu_g$ in the liquid and gas phases, 
respectively.

We choose the well known ``$\phi^4$ theory'' to model the phase separation into two distinct 
phases $\phi = \pm 1$.
This has a pressure tensor and chemical potential given by
\begin{eqnarray}
  P_{\alpha \beta} &=&  \left(p_0
- \kappa \phi \nabla^2 \phi - \frac \kappa 2 |\nabla \phi|^2 \right)  \delta_{\alpha \beta} + 
\kappa \partial_{\alpha} \phi  \, \partial_{\beta} \phi,\\
\mu &=& A\left(- \phi + \phi^3 \right)- \kappa \nabla^2 \phi. 
\label{chempot}
\end{eqnarray}
The bulk pressure in this case is 
\begin{eqnarray}
p_0 = \rho \tfrac{c^2}{3} + A\left(- \tfrac{1}{2} \phi^2 + \tfrac{3}{4} \phi^4 \right).
\end{eqnarray}
In this study, we use the parameters $A = 0.04$ and $\kappa = 0.04$, which lead to an 
interface width of $\sim 4$ lattice sites and a liquid-gas surface tension of $\gamma_{lg} = 
0.0389$. The average density of the system was taken to be $\bar{\rho} = 1$.

As well as a time evolution equation for $f_i$ in Eq. (\ref{latbolt}), which gives the 
Navier-Stokes equation, there is a second lattice Boltzmann equation
\begin{eqnarray}
{\bf g}({\bf r} + {\bf e} \Delta t , t+\Delta t) = {\bf g}^{eq}({\bf r},t) 
\label{latbolt2}
\end{eqnarray}
which describes the time evolution of $g_i({\bf r},t)$, the order parameter distribution 
function. An expression for the equilibrium distribution ${\bf g}^{eq}$ is given in Appendix 
\ref{app1}.
This leads to an advection diffusion equation for the order parameter
\begin{eqnarray}
\partial_t{\phi} + \partial_\alpha \left( \phi u_\alpha \right) = M \nabla^2 \mu,
\label{diff}
\end{eqnarray}
where $M$ is a mobility parameter.

\subsection{Numerical results}

Firstly, the equilibrium properties of the system were verified  \cite{Briant2}. Figure \ref{fig2}(b) shows 
the contact angle obtained numerically as a function of the equilibrium gradient of $\phi$ at 
the boundary. As with the liquid-gas model, good agreement is found with Young's law. (Note 
that the correct curve, in the case when there is a viscosity difference between the liquid 
and gas phases, is not produced if a standard BGK lattice Boltzmann algorithm is used. Details 
of why this is the case are presented in \cite{ourpaper}.) 

%
\begin{figure}
\begin{center}
\includegraphics[width = 5.6in]{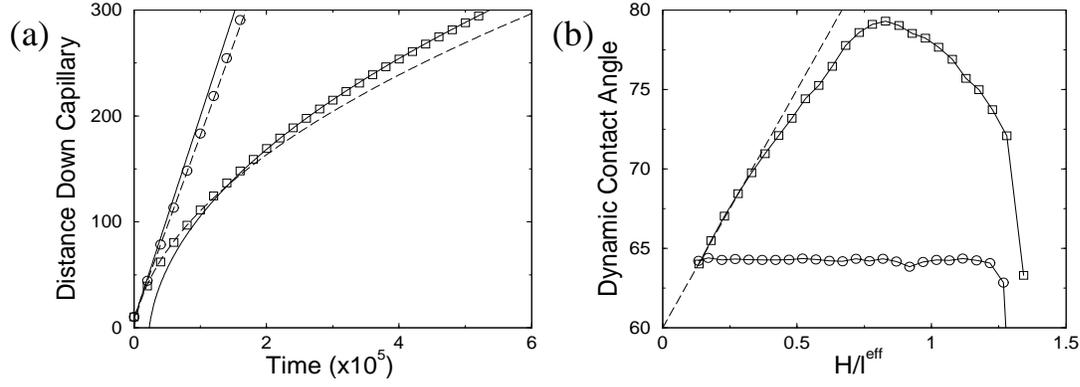}
\end{center}
\caption{(a) The distance of the liquid--gas (A--B) interface along the capillary as a function 
of time for the binary fluid model. Circles: $\nu_l = \nu_g = 1/6$, squares: $\nu_l = 1.17$ 
and $\nu_g = 0.017$, and the solid and dashed lines   are theoretical predictions (see text 
for details). (b) The corresponding variation in dynamic contact angle. The dashed line 
extrapolates the results to the limit $l^{eff} \rightarrow \infty$.}
\label{fig5}
\end{figure}
Figure \ref{fig5}(a) shows a plot of capillary filling distance $l$ as a function of time 
using the binary model. The circles are results when the liquid and gas phases have equal 
viscosity. They form a straight line because viscous dissipation occurs at approximately the 
same rate at any given point down the tube, and so the total dissipation is independent of the 
position of the interface. 
The solid, straight line shows the theoretically expected profile based on Poiseuille flow 
down a tube of length $L=350$ using the numerically measured contact angle of 
$\theta=63^\circ$ (see below). The agreement is not exact as we have ignored dissipation at 
the inlet and outlet of the tube. This can be taken into account by extending the effective 
length of the tube either end by $\simeq H/2$. Subsequently, we use an effective length of 
filling $l^{eff} = l + H/2$, as shown in Fig. \ref{fig7}(b). The dashed, theoretical curve in 
Fig. \ref{fig5}(a) takes into account this refinement and shows much closer agreement with the 
simulation results.

Measurements of the meniscus contact angle $\theta$ (see Fig. \ref{fig7}(b)) were made by 
performing a least squares fit of the interface profile to a circular section. The circles in 
Fig. \ref{fig5}(b) show $\theta$ as a function of $H/l^{eff}$. (Note that $l^{eff}$ increases 
in time so this plot starts on the right side of the diagram and moves left.) The noise in 
this curve is not real but a result of the measurement technique.
We observed that $\theta$ rapidly reaches a stable value of $\theta \simeq 63^\circ$. This is 
significantly different from the equilibrium contact angle of $\theta^{eq} = 60^\circ$. 
\begin{figure}
\begin{center}
\includegraphics[width = 5.6in]{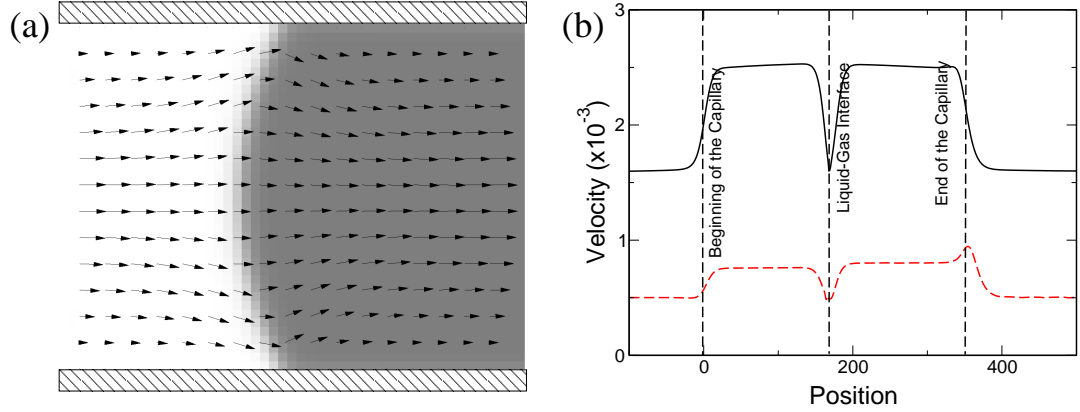}
\end{center}
\caption{(a) The velocity field around the liquid-gas  (A--B) interface during capillary 
filling using the binary 
model ($\nu_l = 0.83$, $\nu_g = 0.067$). (b) The $x$ component of velocity as a function of 
$x$ along the centre of the capillary. Solid line: $\nu_l=\nu_g=1/6$, dashed line: $\nu_l=0.83, 
\nu_g = 0.067$. }
\label{fig4}
\end{figure}
The reason behind this is revealed if we look at the flow field near to the meniscus, as 
plotted by the black arrows in Fig. \ref{fig4}(a). In the bulk phases, the flow down the 
capillary tube is parabolic. On the other hand, the interface itself moves at a constant 
velocity, except very close to the boundaries where the non-slip boundary conditions prevent 
this (in this case diffusion allows the ``slip''). The transition between these two profiles 
is achieved by fluid in the liquid phase being driven into the corners, towards the contact 
points, and, conversely, the gas phase being pushed away (observe the arrows near to the 
contact points in Fig. \ref{fig4}(a)). The driving force for this process is the difference 
between the dynamic and equilibrium contact angles $\Delta \theta = \theta-\theta^{eq}$. 
Another way of looking at this is to say that, because energy is dissipated in the flow field 
near the interface, the dynamic contact angle $\theta$ that appears in the Washburn relation 
(\ref{eq3}) must be greater than $\theta^{eq}$ to reflect the fact that there is less 
available energy to do useful work (in this context useful work means the energy for pulling 
the fluid into the capillary). 

The curves in Fig. \ref{fig4}(b) clearly show that the velocity in the liquid and gas phases 
are identical, as compared to the very different situation that was observed in the liquid-gas 
system in Fig. \ref{fig3}(b). The dip in velocity at the interface is because of the change in 
flow profile, as discussed in the previous paragraph. The dashed curve is for a high viscosity 
ratio, and the bump in the velocity profile at the end of the capillary resulted from the 
formation of vortices indicative of a moderate Reynolds number in the gas phase.

The squares in Fig. \ref{fig5}(a) show the results of a simulation with a large viscosity 
ratio between the two phases of $\nu_l/\nu_g = 70$, sufficiently high that viscous dissipation 
in the gas phase can, justifiably, be ignored. The dashed curve is from Washburn's law based 
on a dynamic contact angle of $\theta = 63^\circ$. Very good agreement is achieved at late 
times. The difference at early times can be explained if we examine the dynamic contact angle 
in Fig. \ref{fig5}(b). In this case, $\Delta \theta$ starts off large and gradually decreases 
as $l^{eff}$ gets bigger. The value $\theta = 63^\circ$ was used above because it is 
representative of the late stages of the simulation. If we set $\theta = 68^\circ$ in 
Washburn's law then we obtain the dotted curve, which can be fitted well to the early time 
behaviour. The dashed line in Fig. \ref{fig5}(b) is a linear extrapolation of the contact 
angle data showing that, as the liquid penetrates far into the tube and the meniscus velocity 
tends to zero, $\theta = \theta^{eq}$ as expected.


\subsection{The role of particle mobility, $M$}

Unlike the liquid-gas model, the binary model has an extra parameter, $M$, which determines 
the diffusive behaviour of the liquid (A) and gas (B) particles.
\begin{figure}
\begin{center}
\includegraphics[width = 5.6in]{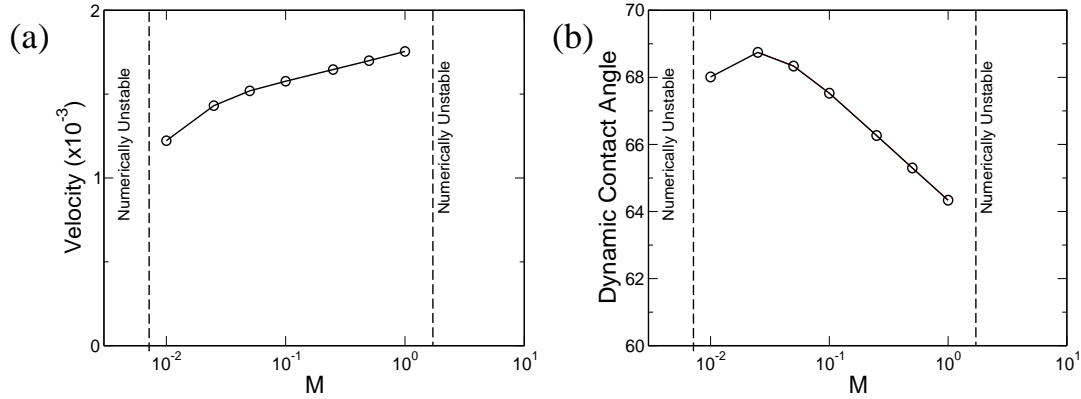}
\end{center}
\caption{(a) The variation in the velocity of capillary filling as a function of mobility $M$, 
using $\nu_l =   \nu_g = 1/6$. (b) The corresponding change in dynamic contact angle.}
\label{fig6}
\end{figure}
To assess the effect of varying $M$ on the system, we consider capillary filling for the case 
when the
the viscosity of the two phases is the same. This particularly simple case was chosen because 
the velocity of filling is constant and can be calculated from a linear fit to the filling 
profile.
Figure \ref{fig6}(a) shows the velocity of filling against $M$. In the limits of small or high 
$M$ the algorithm was found to be numerically unstable. In the stable region there is a 
general trend of increasing velocity with increasing $M$.
This is rather an unexpected result. The right hand side of the diffusion equation 
(\ref{diff}) is normally associated with dissipation of energy towards a free energy minimum 
{(\ref{diff}) and yet we find that increasing its size {\it increases} the energy available to 
do work (in this case work in filling the capillary tube). 
The reason that the contact angle decreases with increasing $M$ is that the diffusion 
alleviates the no-slip boundary conditions -- for larger $M$ the interface slips more easily 
across the surface. Thus dissipation due to flow in the vicinity of the meniscus, and hence 
$\Delta \theta$, are reduced.  


 

\section{Summary}

We have assessed two different ways of simulating capillary flow. Firstly, using a liquid-gas 
approach, the filled length of a capillary tube was found to increase much more rapidly than 
predicted by Washburn's law. This discrepancy was due to condensation of the gas phase at the 
interface. This is allowed within the formalism we used because the van der Waals equation of 
state describes a liquid in equilibrium with its vapour. Moreover, the effect of condensation 
is large because the liquid and gas have similar densities. Essentially we are modelling a 
system close to its critical point. Results in \cite{succi1} show that a good fit to the Washburn 
equation is obtained if the liquid-gas density ratio is large.

We then considered a binary fluid, comprising two different types of particles, where 
evaporation and condensation is not permitted. Very good agreement was achieved with the 
theoretical expression in Eq. (\ref{eq3}), as long as the dynamic contact angle was used in 
the fit. We argued that the dynamic contact angle differed from the equilibrium value because 
of dissipation of energy from the flow field near to the meniscus. 

We hope that this 
simulation technique will prove useful in studying capillary filling in porous media and in 
microchannels with differing geometries or surface structures and patterning.   


\section{Ackowledgements}
We thank S. Succi and L. Biferale for helpful comments. We acknowledge funding from the Office 
of Naval Research, USA and the EU INFLUS project. 

\appendix

\section{Details of the lattice Boltzmann scheme}
\label{app1}

This appendix gives details of the lattice Boltzmann scheme used for this study. 
The matrix ${\bf M}$, that describes a change of basis, is given by
\begin{eqnarray}
{\bf M} = \left(
\begin{array}{ccccccccc}
1 & 1 & 1 & 1 & 1 & 1 & 1 & 1 & 1\\
-4 & -1 & -1 & -1 & -1 & 2 & 2 & 2 & 2\\ 
4 & -2 & -2 & -2 & -2 & 1 & 1 & 1 & 1\\
0 & 1 & -1 & 0 & 0 & 1 & -1 & -1 & 1 \\
0 & -2 & 2 & 0 & 0 & 1 & -1 & -1 & 1 \\
0 & 0 & 0 & 1 & -1 & 1 & -1 & 1 & -1 \\
0 & 0 & 0 & -2 & 2 & 1 & -1 & 1 & -1 \\
0 & 1 & 1 & -1 & -1 & 0 & 0 & 0 & 0 \\
0 & 0 & 0 & 0 & 0 & 1 & 1 & -1 & -1 \\
\end{array}
\right).
\end{eqnarray}
In the new basis the diagonal matrix
\begin{eqnarray}
{\bf S} = \text{diag} \left(0,1,1,0,1,0,1, \omega,\omega \right)
\end{eqnarray}
sets the relaxation rates of different modes. Some of these are arbitrarily set to zero and 
these correspond to conserved quantities, {\it e.g.} the top line dotted with ${\bf f}$ gives 
the density $\rho$. The quantity $\omega = 2/ (6\nu +1)$ sets the kinematic viscosity. Because 
$\nu$ is a function of $\phi$ in Eq. (\ref{nu}) it might seem necessary to calculate the 
collision matrix, ${\bf M}^{-1} {\bf S}{\bf M}$, at each node at each time-step, which would 
be computationally very slow. Our approach is to make a look up table containing $\sim 10^4$ 
matrices with different values of viscosity and simply pick the closest match. 

The equilibrium distribution can be written in the form
\begin{eqnarray}
f^{eq}_i({\bf r}) &=& \tfrac{w_{i}}{c^{2}} \Big( p_0 - \kappa \rho \nabla^2 \rho + e_{i\alpha} \rho 
u_{\alpha} 
+ \tfrac{3}{2c^{2}} \left[ e_{i\alpha} e_{i\beta} - 
\tfrac{c^{2}}{3}\delta_{\alpha\beta}\right] \times \nonumber\\
&& \quad \left( \rho u_\alpha u_\beta + \lambda \left[ u_\alpha \partial_\beta \rho + u_\beta 
\partial_\alpha \rho + \delta_{\alpha \beta} u_\gamma \partial_\gamma \rho \right] \right) 
\Big) \nonumber\\
&&
\hspace{-1cm}
\quad + \tfrac{\kappa}{c^2} \Big( w_i^{xx}  \partial_x \rho \partial_x \rho  + w_i^{yy} 
\partial_y \rho \partial_y \rho + w_i^{xy} \partial_x \rho \partial_y \rho \Big), 
\label{equilibrium}
\end{eqnarray}
for $i=1,..,8$, where $w_{1\text{-}4} = \tfrac{1}{3}$, $w_{5\text{-}8} = \tfrac{1}{12}$, and 
summation over repeated indices is assumed.
Other parameters are $w_{1\text{-}2}^{xx} = w_{3\text{-}4}^{yy} = \tfrac{1}{3}$, 
$w_{3\text{-}4}^{xx} = w_{1\text{-}2}^{yy} = -\tfrac{1}{6}$, $w_{5\text{-}8}^{xx} = 
w_{5\text{-}8}^{yy} = -\tfrac{1}{24}$ and $w_{1\text{-}4}^{xy} = 0$, $w_{5,6}^{xy} = 
\tfrac{1}{4}$, and $w_{7,8}^{xy} = -\tfrac{1}{4}$.
This choice was made to reduce spurious velocities generated at interfaces \cite{kalli}.

The $i=0$ stationary value is chosen to conserve mass:
\begin{equation}
f_0^{eq}({\bf r}) = \rho - \sum_{i=1}^{8} f_i^{eq}({\bf r}).
\label{eqn:feq0}
\end{equation}

For the binary model, the equilibrium distribution $g_i^{eq}$ is given by
\begin{eqnarray}
g^{eq}_i({\bf r}) &=& \tfrac{w_{i}}{c^{2}} \Big( \tfrac{2 M}{\Delta t} \mu  + e_{i\alpha} \phi 
u_{\alpha} + \tfrac{3}{2c^{2}} \left[ e_{i\alpha} e_{i\beta} -
\tfrac{c^{2}}{3}\delta_{\alpha\beta}\right] \phi u_\alpha u_\beta  \Big), \nonumber\\
g_0^{eq}({\bf r}) &=& \phi - \sum_{i=1}^{8} g_i^{eq}({\bf r}).
\end{eqnarray}
For the order parameter distribution functions, the relaxation rates are all set to 1.
During the lattice Boltzmann procedure, it is necessary to numerically calculate both 
derivatives ({\it e.g.} $\partial_x \rho$ in the equilibrium distribution (\ref{equilibrium})) 
and the Laplacian ({\it e.g.} to obtain the chemical potential (\ref{chempot})). These 
continuous quantities are calculated from stencils, discrete operators which use neighbouring 
lattice sites. 
The best choice of stencils to reduce spurious velocities is given by:
\begin{eqnarray}
\bar{\partial}_x  = 
\tfrac{1}{12 \Delta x}
\left[
\begin{array}{ccc}
-1 & 0 & 1 \\
-4 & 0 & 4 \\
-1 & 0 & 1 \\
\end{array}
\right]
, 
\bar{\nabla}^2  = 
\tfrac{1}{6 {\Delta x}^2}
\left[
\begin{array}{ccc}
1 & 4 & 1 \\
4 & -20 & 4 \\
1 & 4 & 1 \\
\end{array}
\right].
\label{best}
\end{eqnarray}

\end{document}